\documentclass[namedreferences]{solarphysics}
\usepackage[hyperref,optionalrh,solaromanenum]{spr-sola-addons} 
\usepackage{epsfig}                     
\usepackage{graphicx}                    
\usepackage{color}                       
\usepackage{breakurl}                         
\usepackage[T1]{fontenc}
\usepackage{amssymb}
\usepackage{tabularx}
\usepackage{multirow}
\usepackage{wasysym}

\newcommand{\etal}{et al.}

\renewcommand{\vec}[1]{{\mathbfit #1}}
\newcommand{\deriv}[2]{\frac{{\mathrm d} #1}{{\mathrm d} #2}}

\newcommand{\grad}{ {\bf \nabla } }
\newcommand{\curl}{ {\bf \nabla} \times}

\chardef\us=`\_

\begin{document}

\begin{article}

\begin{opening}
	
\title{Longitudinal Plasma Motions Generated by Shear Alfv{\'e}n Waves in Plasma with Thermal Misbalance}

%
\author[addressref={aff1,aff2},corref,email={mr\_beloff@mail.ru}]{\inits{S.A.}\fnm{S.}~\lnm{Belov}\orcid{0000-0002-3505-9542}}
\author[addressref={aff3},corref,email={}]{\inits{S.}\fnm{S.}~\lnm{Vasheghani Farahani}\orcid{}}
\author[addressref={aff1,aff2},corref,email={}]{\inits{N.E.}\fnm{N.}~\lnm{Molevich}\orcid{0000-0001-5950-5394}}
\author[addressref={aff1,aff2},corref,email={}]{\inits{D.I.}\fnm{D.}~\lnm{Zavershinskii}\orcid{0000-0002-3746-7064}}

%

\runningauthor{S. Belov \etal}
\runningtitle{Longitudinal Plasma Motions Generated by Shear Alfv{\'e}n Waves}

\address[id=aff1]{Department of Physics, Samara National Research University, Moscovskoe sh. 34, Samara, 443086, Russia}
\address[id=aff2]{Department of Theoretical Physics, Lebedev Physical Institute, Novo-Sadovaya st. 221, Samara, 443011, Russia}
\address[id=aff3]{Department of Physics, Tafresh University, Tafresh 39518 79611, Iran}

\begin{abstract}
Compressional plasma perturbations may cause thermal misbalance between plasma heating and cooling processes. This misbalance significantly affects the dispersion properties of compressional waves providing a feedback between the perturbations and plasmas. It has been shown that Alfv{\'e}n waves may induce longitudinal (compressional) plasma motions. In the present study, we analyze the effects of thermal misbalance caused by longitudinal plasma motions induced by shear Alfv{\'e}n waves. We show that thermal misbalance leads to appearance of exponential bulk flows, which itself modifies the Alfv{\'e}n-induced plasma motions. In the case of sinusoidal Alfv{\'e}n waves, we show how the amplitude and phase shift of induced longitudinal motions gain dependence on the Alfv{\'e}n wave frequency while shedding light on its functionality. This feature has been investigated analytically in application to coronal conditions. We also consider the evolution of longitudinal plasma motions induced by the shear sinusoidal Alfv{\'e}n wave  by numerical methods before comparing the results obtained with our presented analytical predictions to justify the model under consideration in the present study.
\end{abstract}


\end{opening}

\section{Introduction}
A perfect domain in physics to study the character of Alfv\'en waves together with their seismological aspects is the Sun. This statement is not just supported by the knowledge that the Alfv\'en wave is the primary candidate for carrying energy from the lower to outer solar atmosphere, but benefits from the Alfv\'en-wave contribution towards understanding coronal heating. An important feature of the Alfv\'en wave that is of interest in the present study is rooted in its nature since its presence is associated with the magnetic-tension force \citep{1942Natur.150..405A}. 

The magnetic-tension force is one of the main nonlinear forces connected with the Alfv\'en wave \citep{2011A&A...526A..80V} that affects the compressive perturbations in a plasma medium. This is in the sense that the interplay of these forces in a transversely structured plasma structure provides cylindrical collimation through the compressive perturbations, especially the longitudinal perturbations, owing to the Bernoulli effect \citep{2017ApJ...844..148V}. A prominent nonlinear force in this regard is the ponderomotive force \citep{1999JPlPh..62..219V}, which in specific ideal conditions acts as the only force that sustains jet collimation \citep{2018ApJ...869...93M}. Even in the absence of transverse structuring in a uniform plasma medium that hosts a magnetic null-point \citep{2019A&A...624A..90P}, the ponderomotive force connected with the Alfv\'en wave induces longitudinal, together with transverse magnetoacoustic, waves that contribute towards energy transfer in the solar atmosphere \citep{2018MNRAS.479.4991S}. The back reaction of the density perturbations together with the longitudinal induced flows initiated by the ponderomotive force on the initially propagating Alfv\'en wave creates a non-uniform background in the medium \citep{2020NatSR..1015603S}. This non-uniformity would act as if the plasma were initially inhomogeneous, which results in a famous physical concept known as phase mixing \citep{2013A&A...555A..86T}, which itself is a damping mechanism \citep{1997SoPh..175...93N,2005A&A...440..357C,2018A&A...620A..44R}. Regarding a visco-resistive medium with inhomogeneous plasma density, the induced longitudinal flows due to the ponderomotive force provide local temperature increases that trigger longitudinal flows in the context of phase mixing \citep{McLaughlin2011, 2017ApJ...840...64S}. 

The solar corona makes the interplay between magnetic and compressional perturbations even more intriguing. Temperature and/or density variations caused by magnetoacoustic (MA) waves or induced by magnetic perturbations disturb the balance between plasma heating and radiative-cooling rates allowing the thermal-misbalance effect \citep{Field1965, Molevich88, 2017ApJ...849...62N} to develop. The origin of thermal misbalance comes from the  dependence of coronal heating together with the cooling rates on plasma parameters such as density and temperature. The changes of rates due to compression affect the variations, which results in re-establishing the interplay between plasma-heating/cooling processes and compressional perturbation. Thus, the interplay between magnetic perturbations/forces and non-adiabatic processes via compression variations can be established as well. This non-linear effect is the subject of the current research.

The thermal misbalance significantly affects the dispersion properties of compressional eigenmode leading to mode amplification/damping, frequency dependence of the phase speed, and increment/decrement \citep{ Zavershinskii2019, Molevich88} and a number of effects visible on either linear and non-linear stages of wave evolution \citep[see][etc.]{  2010PhPl...17c2107C,2017ApJ...849...62N,2019A&A...624A..96C, Zavershinskii2020, 2020PPCF...62a4016A}. Interestingly, the temporal scales determining properties of compressional perturbations are the scales in which the cooling/heating misbalance function depends on the plasma parameters \citep[see][for details]{Zavershinskii2019} not the cooling or heating timescales. It has been shown that these timescales are of the same order as observed MA waves and oscillations \citep{Zavershinskii2019, 2019A&A...628A.133K}. Analysis conducted by \cite{2019A&A...628A.133K} showed that a reasonable choice of the unspecified coronal-heating function allows one to reproduce the observed properties of standing slow MA oscillations. Subsequently, \cite{Kolotkov_2020} seismologically constrained the coronal-heating function using observations of slow MA waves in various coronal-plasma structures. Considering some specific mechanisms satisfying seismological constaints,  \cite{2021SoPh..296...20P} revealed that inclusion of thermal misbalance in addition to thermal conductivity and compressive viscosity better matches the predicted scaling law between the damping time and wave period for the observed \textit{ Solar Ultraviolet Measurements of Emitted Radiation} (SUMER) oscillations. Regarding the propagation of Alfv\'en waves in a non-ideal plasma medium that resembles coronal holes, the thermal misbalance modifies the wave-steepening scale so that for lower solar atmospheric conditions the Alfv\'en-wave steepening is strengthened while experiencing either forward or backward shock. However, as for the slow magnetoacoustic waves where either amplification or damping is observed \citep{2017ApJ...849...62N}, the thermal misbalance may either steepen or dampen the amplitude of the Alfv\'en wave subject to conditions \citep{Belov_2019,2020SoPh..295..160B}. It is worth stating that Alfv\'en waves may be bi-exponentially amplified due to parametric interactions with isentropically unstable acoustic/magnetoacoustic waves \citep{2015Ap&SS.358...22Z, Belov2018, Belov2019RuPhJ}. However, when transverse structuring exists, torsional Alfv{\'e}n waves are possible candidates for carrying vast amounts of energy in the lower solar atmosphere contributing towards coronal heating \citep{2017NatSR...743147S}. This is why when transverse structuring is present in the plasma medium, the Alfv\'en-wave steepening would eventually form a shock that is directly proportional to the equilibrium flow speed that provides a shear at the boundary of the plasma structure \citep{2021ApJ...906...70F}. In addition to shocks, pseudo-shocks also contribute towards energy supply to the inner solar corona providing localized mass transport together with coronal heating \citep{2018NatAs...2..951S}.
 
In the present study, the aim is to shed light on the perturbations induced by shear Alfv\'en waves in the presence of thermal misbalance by focusing on the nonlinear forces connected with the shear Alfv\'en wave, where in the succeeding section the model and equilibrium conditions are to be presented. In the third section, expressions governing the induced perturbations are provided before carrying out parametric studies regarding the scaling behavior of the amplitude and phase shift as well as comparison between the analytic solutions and results obtained numerically. In the final section, the conclusions are summarized.

\section{Model and Equilibrium Conditions}\label{s:Model} 

Consider a fully ionized plasma medium that resembles the solar atmosphere, where due to the macroscopic view of the present study it provides a perfect domain for implementing the MHD theory. In this line the MHD equations are taken under consideration to model the dynamics of Alfv\'en waves in an initially homogeneous plasma medium that experiences a misbalance between non-adiabatic heating and radiative losses \citep{Priest2014}. The MHD set of equations appropriate for the aims of the present study are expressed as

\begin{equation}
\frac{\partial{\rho}}{\partial{t}}+\nabla\cdot\left(\rho\vec{v}\right)=0\,,
\label{Cont}
\end{equation}
\begin{equation}
\rho\left(\frac{\partial{\vec{v}}}{\partial{t}}
+\left(\vec{v}\vec{\cdot}\vec{\nabla}\right)\vec{v}\right)
=-\vec{\grad}\!P-\frac{1}{4\pi}\vec{B}\times\left(\curl\vec{B}\right),
\label{Motion}
\end{equation}
\begin{equation}
\frac{\partial{\vec{B}}}{\partial{t}}=\curl\left(\vec{v}\times\vec{B}\right),
\label{Induction}
\end{equation}
\begin{eqnarray}
\nabla\cdot\vec{B}=0 \,,
\label{Div}
\end{eqnarray}
\begin{equation}
C_{V\!\infty}\rho\left(\frac{\partial{T}}{\partial{t}}
+\left(\vec{v}\vec{\cdot}\vec{\nabla}\right)T\right)-
\frac{k_{\mathrm{B}}T}{m}\left(\frac{\partial{\rho}}{\partial{t}}
+\left(\vec{v}\vec{\cdot}\vec{\nabla}\right)\rho\right)=-\rho\,\!Q\!\left(\rho, T\right),
\label{Energy}
\end{equation}
\begin{equation}
P=\frac{k_\mathrm{B}}{m}\rho T\,,
\label{State}
\end{equation}
where $\rho$, $T$, and $P$ respectively represent the density, temperature, and pressure of the plasma, while $\vec{v}$ and $\vec{B}$ are vectors of the plasma velocity and magnetic field. The Boltzmann constant, the mean mass per volume, and the high-frequency specific heat capacity at constant volume are respectively shown by $k_\mathrm{B}$, $m$, and $C_{V\!\infty}$. The term  $Q\!\left(\rho, T\right)=L\!\left(\rho, T\right)-H\!\left(\rho, T\right)$ is the generalized heat-loss function (\citealp{Parker1953, Field1965}) expressed by the cooling and heating rates represented by $L\!\left(\rho, T\right)$ and $H\!\left(\rho, T\right)$, respectively. It is worth noting that the heat-loss function  $Q\!\left(\rho, T\right) $ equals zero under steady-state conditions where we have $Q\!\left(\rho_0, T_0\right)=L\!\left(\rho_0,T_0\right)-H\!\left(\rho_0, T_0\right)=L_0-H_0=0$. As the cooling and heating rates depend on density and temperature, the perturbations of the physical variables initiate the thermal misbalance where the reaction on the wave dynamics is of interest in the present study. Further, we consider a propagating 1D linear polarized shear Alfv{\'e}n wave polarized in the $x$-direction propagating along the $z$-axis in the positive direction parallel to the external magnetic field $B_z=B_0$. In the weakly nonlinear regime, the perturbation theory up to the second order of small parameter $\alpha\ll1$, $\rho=\rho_0+\alpha\,\rho_1+\alpha^2\!\rho_2$ and $B_x=\alpha\,B_{x1}+\alpha^2\!B_{x2}$ is applied to the set of equations expressed by Equations \ref{Cont}\,--\,\ref{State}. By assuming that there are no acoustic perturbations in the first order of $\alpha$, Equations \ref{Cont}\,--\,\ref{State} could be simplified to 
\begin{eqnarray}
\frac{\partial{}}{\partial{t}}\left(\frac{\partial^2{v_{z}}}{\partial{t}^2}-
C_\infty^2\frac{\partial^2{v_{z}}}{\partial{z}^2}+\frac{\partial^2}{\partial{z}\partial{t}}\frac{B_{x}^2}{8\pi\!\rho_0}\right)+\nonumber\\
+\frac{1}{\tau_V}\left(\frac{\partial^2{v_{z}}}{\partial{t}^2}-
C_0^2\frac{\partial^2{v_{z}}}{\partial{z}^2}+\frac{\partial^2}{\partial{z}\partial{t}}\frac{B_{x}^2}{8\pi\!\rho_0}\right)=0\,.
\label{AcousticWaveEq}
\end{eqnarray}
\begin{equation}
\frac{\partial^2{\!B_{x}}}{\partial{t}^2}-
C_\mathrm{A}^2\frac{\partial^2{\!B_{x}}}{\partial{z}^2}=0,
\label{AlfvenWaveEq}
\end{equation}
where $C_\infty^2=C_{P\infty}k_B\!T_0/C_{V\infty}m$ is the square of high-frequency ($\omega \left|\tau_V\right|\gg1$) sound speed (the standard value of the sound speed in the medium without the thermal misbalance) and $C_0^2=\left(Q_{0T}-Q_{0\rho}\rho_0/T_0\right)k_B\!T_0$ $/Q_{0T}m$ is the square of the low-frequency ($\omega\left|\tau_V\right|\ll1$) sound speed that arises due to the thermal misbalance. The sound speed in the medium under consideration is frequency dependent and varies between $C_0$ and $C_\infty$ \citep[see, e.g.,][]{Molevich88,Zavershinskii2019}. It is worth stating that in case of non-misbalance conditions, there is no dispersion caused by the heating/cooling processes and acoustic waves propagate with the phase speed $C_\infty$. Here, also, $C_\mathrm{A}^2=B_0^2/4\pi\rho_0$ is the square of Alfv{\'e}n speed; $C_{P\infty}=C_{V\infty}+k_\mathrm{B}/m$ is the high-frequency specific heat capacity at constant pressure; $\tau_V=C_{V\infty}/Q_{0T}$ is the characteristic time associated with the thermal misbalance \citep[see $\tau_2   $ in][]{Zavershinskii2019,Kolotkov_2020};  $Q_{0T}=\left.\partial Q/\partial T\right|_{\rho_0, T_0}$, $Q_{0\rho}=\left.\partial Q/\partial\rho\right|_{\rho_0, T_0}$. {It should be mentioned that $v_z \propto \alpha^2$ (second-order perturbation), while $B_x \propto \alpha$ (first-order perturbation), so that Equation \ref{AcousticWaveEq} describes second-order effects.}

It can be seen from Equation \ref{AcousticWaveEq}, that the Alfv{\'e}n wave induces longitudinal plasma motions due to the magnetic pressure $\left[B_x^2/8\pi\right]$ gradient. At the same time, Equation \ref{AlfvenWaveEq} has the exact solution in the form of an Alfv{\'e}n wave propagating in the positive direction in the form of $B_x=\Phi\left(\xi\right)$, where we have $\xi=z-C_\mathrm{A}t$. Looking for the Alfv{\'e}n-induced motion, we look for the solution of Equation \ref{AcousticWaveEq} of the form $v_z=v_z\left(\xi\right)$. In this case, Equation \ref{AcousticWaveEq} may be written as a simple ordinary differential equation  
\begin{equation}
\deriv{}{\xi}\left(C_{\mathrm{A}}\left(C_\infty^2-C_\mathrm{A}^2\right)v_z+\frac{C_\mathrm{A}^4}{2 B_0^2}B_x^2\right)+\frac{1}{\tau_V}\left(\left(C_\mathrm{A}^2-C_0^2\right)v_z-\frac{C_\mathrm{A}^3}{2 B_0^2}B_x^2\right)=0.
\label{AcousticWaveEqProfile}
\end{equation}
Equation \ref{AcousticWaveEqProfile} allows us to express $v_z$ as a function of $B_x^2$. This solution can be seen from Equation \ref{AcousticInt}, where $C$ is an arbitrary constant determined from the condition $\left. v_z\right|_{\xi=0}=0$. Thus, we have 
\begin{eqnarray}
v_{z}=C \mathrm{e}^{\Psi\xi}+K_{\!1}B_{x}^2
+K_2\mathrm{e}^{\Psi\xi}\int \mathrm{e}^{-\Psi\xi}B_{x}^2\,\mathrm{d}\xi,\qquad\qquad\qquad
\label{AcousticInt}\\
K_{\!1}=\frac{C_\mathrm{A}^3}{2\!\left(C_\mathrm{A}^2-C_\infty^2\right)\!B_0^2},\quad
K_{\!2}=\frac{C_\mathrm{A}^2\left(C_\infty^2-C_0^2\right)}{2\tau_V\!\left(C_\mathrm{A}^2-C_\infty^2\right)^2\!B_0^2},\quad
\Psi=\frac{\left(C_\mathrm{A}^2-C_0^2\right)}{C_\mathrm{A}\tau_V\!\left(C_\mathrm{A}^2-C_\infty^2\right)}.\nonumber
\end{eqnarray}
Equation \ref{AcousticInt} determines longitudinal plasma motions induced by Alfv{\'e}n waves in a plasma with thermal misbalance. It is worth noting that $K_1$ and $K_2$ should be of the order unity to satisfy the expansion order, {which implies, in particular, that $\beta<1$.}

 The first term of the solution expressed by Equation \ref{AcousticInt} is a bulk flow analogous to the bulk flow obtained by \cite{McLaughlin2011} for the case of visco-resistive plasma. This flow arises due to the fact that thermal misbalance causes local increase/decrease in temperature, which leads to an increase in the  pressure gradients that drives the bulk flow. We should stress specific features following from the consideration of thermal misbalance. In the visco-resistive case, the flow follows the direction of the Alfv\'en wave. But, in thermally active plasmas, the flow travels either in the direction of the Alfv\'en wave or opposite to it. In the following section, we will introduce the conditions for same- and opposite- direction flows considering the sinusoidal inducing Alfv\'en wave.  
 
 The second term of the solution expressed by Equation \ref{AcousticInt} coincides with the Alfv{\'e}n-induced motion in an ideal plasma without thermal misbalance. The last term of the solution expressed by Equation \ref{AcousticInt} describes how the thermal misbalance affects the longitudinal plasma motion induced by the Alfv{\'e}n wave. This influence is as follows: compressional plasma perturbations induced by the magnetic-pressure gradients in the Alfv{\'e}n wave cause the thermal misbalance, which leads to the additional plasma heating or cooling, similar to the case of bulk flow; see the first term. Subsequently, this disturbed heating/cooling introduces gas-dynamic pressure gradients driving longitudinal-plasma motions. A simple mechanical analogue of this process is driven oscillations of a pendulum with a positive/negative friction. The external force with its own frequency moves a pendulum, but a positive/negative friction leads to the energy dissipation/gain that causes the oscillations to experience a phase-shift. Following this analogy, plasma particles act like a pendulum, the  ponderomotive force induced by the Alfv{\'e}n wave is an external force, and the thermal misbalance acts as some positive/negative friction.

\section{Results and Discussions}\label{s:Longitudinal} 

Let us examine the solution expressed by Equation \ref{AcousticInt} for the sinusoidal Alfv{\'e}n wave with relative amplitude $\epsilon\ll1$:
\begin{equation}
B_x=\epsilon B_0 \sin\left(k\xi\right)=\epsilon B_0 \frac{\mathrm{e}^{\mathrm{i}k\xi}-\mathrm{e}^{-\mathrm{i}k\xi}}{2\mathrm{i}}.
\label{AlfvenSol}
\end{equation}
Substitute the signal expressed by Equation \ref{AlfvenSol} into the expression expressed by Equation \ref{AcousticInt}) to obtain Equation \ref{AcousticSol} for the longitudinal plasma velocity with the initial condition $\left. v_z\right|_{\xi=0}=0$ as
\begin{equation}
v_z =C\mathrm{e}^{\Psi\xi} +V_0-\left(A_1\cos\left(2 k\xi\right)-A_2\sin\left(2 k\xi\right)\right)\,,
\label{AcousticSol}
\end{equation}
where
\begin{eqnarray}
C = \frac{\epsilon^2 C_\mathrm{A}^3}{4}\frac{4\omega^2\tau_V^2\left(C_\infty^2-C_0^2\right)\left(C_\mathrm{A}^2-C_{\infty}^2\right)}
{\left(\left(C_\mathrm{A}^2-C_0^2\right)^2+4\omega^2\tau_V^2\left(C_\mathrm{A}^2-C_{\infty}^2\right)^2\right)\left(C_\mathrm{A}^2-C_{0}^2\right)},\nonumber\\
V_0= \frac{\epsilon^2}{4}\frac{C_\mathrm{A}^3}{C_\mathrm{A}^2-C_0^2}\,,\qquad\qquad\qquad\nonumber\\
A_1=\frac{\epsilon^2 C_\mathrm{A}^3}{4}\frac{\left(\left(C_\mathrm{A}^2-C_0^2\right)+4\omega^2\tau_V^2\left(C_\mathrm{A}^2-C_{\infty}^2\right)\right)}
{\left(C_\mathrm{A}^2-C_0^2\right)^2+4\omega^2\tau_V^2\left(C_\mathrm{A}^2-C_{\infty}^2\right)^2}\,,\nonumber\\
A_2=\frac{\epsilon^2C_\mathrm{A}^3}{4}\frac{2\omega\tau_V\left(C_0^2-C_{\infty}^2\right)}
{\left(C_\mathrm{A}^2-C_0^2\right)^2+4\omega^2\tau_V^2\left(C_\mathrm{A}^2-C_{\infty}^2\right)^2},\,\nonumber
\end{eqnarray}
where we have $\omega=C_\mathrm{A} k$. Equation \ref{AcousticSol} may be rewritten in a more convenient form by transformation of the oscillating part as 
\begin{eqnarray}
v_z =C\mathrm{e}^{\Psi\xi} +V_0-A\cos\left(2k\xi+\phi_0\right),\label{AcousticSol2}\\
A = \sqrt{A_1^2+A_2^2}, \qquad  \phi_0=\mathrm{arctan}\left(\frac{A_2}{A_1}\right).\nonumber 
\end{eqnarray}
In the case without thermal misbalance, the RHS of Equation \ref{AcousticSol2} equals $\epsilon^2 C_\mathrm{A}^3\cdot$ $\cdot\left(1-\cos\left(2k\xi\right)\right)/4\left(C_\mathrm{A}^2-C_{\infty}^2\right)$ which coincides with the expression obtained by \cite{McLaughlin2011} for the case of an ideal plasma. The primary feature of the obtained longitudinal motion is the presence of an exponential bulk flow. The direction of this flow is determined by the sign of $C$ (see Equation \ref{AcousticSol}). When we have $C>0$, the flow is co-directed with the Alfv\'en wave, and conversely, if we have $C<0$, the flow is oppositely-directed. For $C_\mathrm{A}^2>C_\infty^2,C_0^2$ (Alfv\'en waves are faster than acoustic waves), this flow is co-directed if $C_\infty^2>C_0^2$. For $\tau_V>0$, this condition coincides with the condition of isentropic stability $\tau_V\left(C_\infty^2-C_0^2\right)>0$ \citep[see, e.g.,][]{Molevich88,Zavershinskii2020}. Thus, the flow is co-directed with the Alfv\'en wave in the isentropically stable plasma and oppositely-directed in the isentropically unstable plasma. Note that this direction for flows is similar to those described earlier in the heat-releasing gaseous media when acoustic flows arose against the propagation of the initiating wave in the case of isentropic instability and along the propagation of the initiating wave in the case of stability \citep{Molevich2001excitation,Molevich2002nonstationary}. The second feature is that the velocity amplitude $A$ of the oscillating part of Equation \ref{AcousticSol2} becomes frequency-dependent. The limiting values of the amplitude are

\begin{equation} \label{amplimits}
	A = \left\{  \begin{array}{ll} A_\mathrm{hf} = \epsilon^2 C_\mathrm{A}^3/4\left(C_\mathrm{A}^2-C_{\infty}^2\right) & \textrm{for ~~} \omega\left|\tau_V\right|\gg1  \textrm{,}\\ A_\mathrm{lf} =  \epsilon^2 C_\mathrm{A}^3/4\left(C_\mathrm{A}^2-C_0^2\right) & \textrm{for ~~} \omega\left|\tau_V\right|\ll1\textrm{.} \end{array} \right.
\end{equation} 

In the isentropically stable medium, where $C_{\infty}^2>C_0^2$, there will be monotonic decay of the velocity amplitude $A$ with respect to period from $ A_\mathrm{hf}$ to $ A_\mathrm{lf}$. Such dependence will lead to a better generation of longitudinal oscillations with shorter periods. In the case of isentropic instability, there will be monotonic growth from $ A_\mathrm{hf}$ to  $ A_\mathrm{lf}$.

The last, but not the least, feature is appearance of the frequency-dependency velocity phase shift $\phi_0$. The maxima of the induced flow is no longer in phase with the Alfv\'en wave maxima. The maximum induced perturbation overtakes the Alfv\'en wave maximum in the isentropically stable plasma and falls behind in the isentropically unstable plasma. As a result, the Alfv\'en wave steepening is modified. The greatest phase-shift is for the Alfv\'en wave period $P_\mathrm{max}=4\pi\tau_V\sqrt{\left(C_\mathrm{A}^2-C_{\infty}^2\right)/\left(C_\mathrm{A}^2-C_0^2\right)}$. Interestingly, in the case of low plasma-$\beta$, where we have $C_\mathrm{A}^2\gg C_{\infty}^2,C_0^2$, the solution expressed by Equation \ref{AcousticSol2} transfers to the solution for ideal plasma conditions \citep{McLaughlin2011}.

\subsection{Frequency Dependence of Amplitude and Phase Shift}\label{Dependency} 

In order to illustrate the frequency dependence of the velocity amplitude and phase shift, we use $n_0=10^{10}\,\mathrm{cm}^{-3}$, $T_0=10^6\,\mathrm{K}$, $B_0=7\,\mathrm{G}$. Also, we assume that the relative amplitude of the Alfv{\'e}n wave is $\epsilon=0.01$. In this study, the loss term due to the optically thin radiation is taken as of the form 
\begin{equation}
\label{loss_f}
L\!\left(\rho,T\right) = \frac{\rho}{4 m^2}\,\Lambda\!\left(T\right)\,,
\end{equation}
where $m =0.6\times 1.67\times10^{-24}\,\mathrm{g}$ is the mean particle mass and $\Lambda\!\left(T\right)$ is the radiative loss function determined from the CHIANTI atomic database v. 10.0 (\citealp{Dere1997, 2021CHIANTI}). The heating function $H\!\left(\rho, T\right)$ can be locally modeled as
\begin{equation}
\label{heat_f}
H\!\left(\rho, T\right)=h\rho^aT^b\,,
\end{equation}
where $h$, $a$, and $b$ are constants. The first constant $h$ is determined from the steady-state condition $Q\!\left(\rho_0,T_0\right)=0$: $h=L\!\left(\rho_0,T_0\right)/\rho_0^a T_0^b$. The power-law indices $a$ and $b$ are associated with a specific heating mechanism. In the present study we follow \cite{Kolotkov_2020} and \cite{ Duckenfield2020effect} and use the values of $a=1/2$, $b=-7/2$, for which thermal misbalance leads to additional damping of all compressional eigenmodes.

\begin{figure}    
	\centerline{\includegraphics[width=.95\textwidth,clip=]{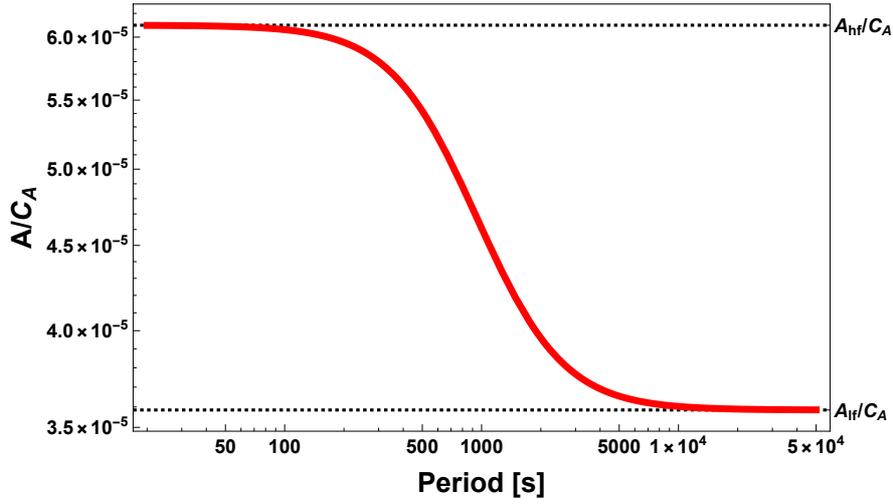}
	}
	\caption{Dependence of the relative amplitude of the oscillating part connected with longitudinal motions induced by the Alfv{\'e}n wave on the Alfv{\'e}n wave period for $T_0=1\,\mathrm{MK}$, $n_0=10^{10}\,\mathrm{cm}^{-3}$ and $B_0=7\,\mathrm{G}$. 
	}
	\label{fig-1}
\end{figure}

\begin{figure}    
	\centerline{\includegraphics[width=.85\textwidth,clip=]{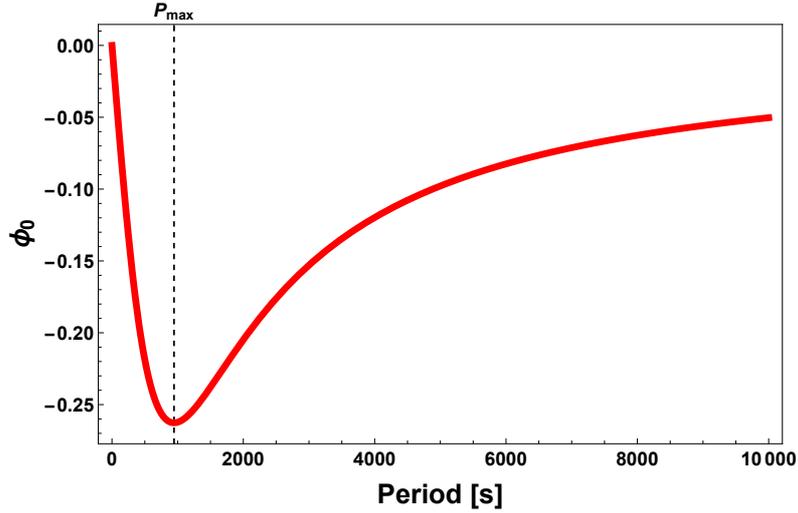}
	}
	\caption{Dependence of the phase shift {(with respect to the case of ideal plasma}) of the oscillating part connected with longitudinal motions induced by the Alfv{\'e}n wave on the Alfv{\'e}n wave period for $T_0=1\,\mathrm{MK}$, $n_0=10^{10}\,\mathrm{cm}^{-3}$ and $B_0=7\,\mathrm{G}$.
	}
	\label{fig-2}
\end{figure}

Figure \ref{fig-1} demonstrates the dependence of the relative amplitude $\left[A/C_\mathrm{A}\right]$ of the oscillating part of the longitudinal motion on the Alfv{\'e}n wave period for the plasma parameters considered. It can be clearly seen that the amplitude monotonically decreases with period. Nevertheless, the absolute value of the  difference between high-frequency and low-frequency amplitude limits $ A_\mathrm{hf}$ and  $ A_\mathrm{lf}$ (Equation \ref{amplimits}) is small. This is primarily due to the choice for the relative amplitude of the Alfv{\'e}n wave $\left[\epsilon=0.01\right]$ and also due to consideration of the previously mentioned heating mechanism $\left[ H\!\left(\rho, T\right) \approx \rho^{\frac{1}{2}}T^{-\frac{7}{2}}\ \right]$. The greater difference between  $ A_\mathrm{hf}$ and  $ A_\mathrm{lf}$ can be reached first by choosing of  the higher relative amplitude $\epsilon$. Furthermore, one may also consider alternative heating mechanisms contributing to greater dispersion (greater difference between $C_0^2$ and $C_\infty^2$ ) and also the different values of magnetic field, which will lead to changes in denominators of $ A_\mathrm{hf}$ and  $ A_\mathrm{lf}$  (Equation \ref{amplimits}).  Figure \ref{fig-2} illustrates dependence of the phase shift $\left[\phi_0\right]$ of the oscillating part of longitudinal motions on the Alfv{\'e}n wave period. The oscillating part of the considered case is the phase shift with maximum shift at $P_\mathrm{max}=945$ seconds {for the chosen heat-loss model and specific combination of the plasma parameters. Variation in heat-loss model will lead to the variation in $P_\mathrm{max}$.} 

\subsection{Comparison with Numerical Simulations}\label{Comparison}   

In order to validate the analytical conclusions regarding the thermal misbalance influence on longitudinal plasma motions induced by Alfv{\'e}n waves,  Equations \ref{Cont}\,--\,\ref{State} have been solved numerically in 1D using the flux-corrected transport technique \citep{Boris73, Toth96}. An Alfv\'en wave has been generated at the left boundary of the numerical domain in the form
\begin{eqnarray}
B_x=\epsilon\!B_0\sin\left(\frac{2\pi}{P}t\right), \qquad v_x=-\frac{B_x}{\sqrt{4\pi\rho_0}}.\nonumber
\end{eqnarray}
We have chosen $\epsilon=0.01$ and $P=300$ seconds for the simulations. Other variables have been kept at their initial values at the left and right boundaries of the numerical domain.

\begin{figure}    
	\centerline{\includegraphics[width=.95\textwidth,clip=]{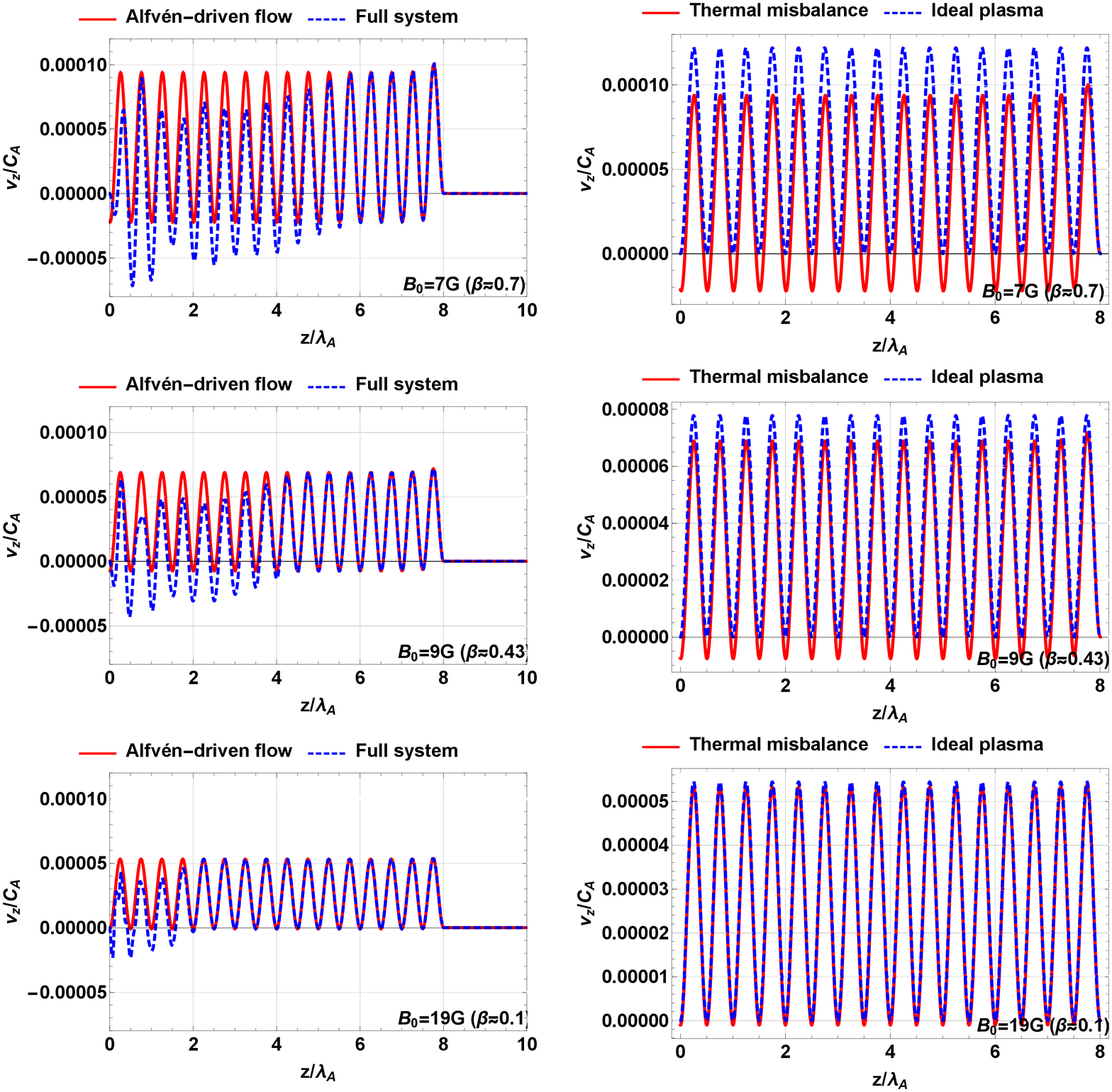}
	}
	\caption{Relative longitudinal plasma velocity for $T_0=1\,\mathrm{MK}$, $n_0=10^{10}\,\mathrm{cm}^{-3}$, and $B_0=7\,\mathrm{G}, 9\,\mathrm{G}$, $19\,\mathrm{G}$. \textit{Left panel}, comparison between the analytical solution for Alfv{\'e}n-induced longitudinal motions (Equation \ref{AcousticSol2}) and numerical solution of the full MHD system (Equations \ref{Cont}\,--\,\ref{State}) in 1D. \textit{Right panel}, comparison between analytical solutions  for Alfv{\'e}n-induced longitudinal motions in plasma with thermal misbalance and ideal plasma conditions.
	}
	\label{fig-3}
\end{figure}

The left panel of Figure \ref{fig-3} presents a comparison between the analytical solution for Alfv{\'e}n-induced longitudinal motions obtained by Equation \ref{AcousticSol2} and the numerical solution of the full 1D-MHD system obtained from Equations \ref{Cont}\,--\,\ref{State}. It can be noticed that the numerical solution regarding the full system is constituted by two parts: acoustic perturbations and Alfv\'en-driven compressional perturbations. These perturbations are analytically described for plasmas without thermal misbalance \citep{McLaughlin2011}. When the Alfv\'en speed exceeds the sound speed, two domains are observed. In the right domain, only the Alfv\'en-driven perturbations occur which are obtained from the analytical solution expressed by Equation \ref{AcousticSol2} that coincides well with the numerical solution. In the left domain, the acoustic and Alfv\'en-driven compressional perturbations are both present, which enables them to interact with each other. It should be mentioned that in contrast to the driven perturbation, the acoustic perturbation  decays due to the chosen heating mechanism $ H\!\left(\rho, T\right) \sim \rho^{\frac{1}{2}}T^{-\frac{7}{2}}\ $. Due to the simultaneous presence of the acoustic and Alfv\'en-driven  perturbations, the numerical solution in this domain deviates from the analytical solution expressed by Equation \ref{AcousticSol2} which describes Alfv\'en-driven  perturbations only. The right panels of Figure \ref{fig-3} compare analytical solutions for Alfv{\'e}n-induced longitudinal motions in plasmas with the thermal misbalance and in ideal plasma conditions.  Firstly, it can be noticed that the solutions almost coincide for the low plasma-$\beta$ conditions. It could also be noticed that the  peaks of the curves 'float up' when thermal misbalance is present. This could be explained due to the appearance of exponential bulk flows, see the first term of Equation \ref{AcousticSol2}. {Because of the heat-loss model and plasma parameters considered here, this flow decays rapidly from the wave front that is why only one peak is noticeably seen to 'float up'.} The 'floating up' signatures become weaker for lower plasma-$\beta$ values. Another aspect of longitudinal motions when thermal misbalance comes in to play is the formation of areas with negative longitudinal plasma speeds. The reason  for this is that the absolute value of  the amplitude $\left[A\right]$ in Equation \ref{AcousticSol2} is greater than the absolute value of $V_0$. This is while a smaller amplitude together with a smaller negative phase shift is  observed, which is presented  by comparison with the oscillating part of the motion in ideal plasma conditions, see Figure \ref{fig-3}. It is worth noting that this feature disappears for lower plasma-$\beta$ conditions. The seismological aspect of the right panels of Figure \ref{fig-3}  is associated with a change in longitudinal velocity in the presence of thermal misbalance,  which is over $30$ percent of the velocity amplitude for plasma-$\beta$ values around $0.7$ which corresponds to the solar chromosphere and corona. This could be a new seismological tool for making constraints on the solar coronal heating function using the line-of-sight velocity observations like one made by \citet{2009A&A...501L..15B} where line-of-sight velocity increase of about $50\,\%$  was observed for coronal holes.

\section{Conclusions}\label{s:Discussion}

In this study, the effects connected with the thermal misbalance caused by longitudinal plasma motions induced by shear Alfv\'en waves have been analyzed. By implementing the MHD set of equations and by considering the misbalance between the energy gain and losses in the energy, of, state equation, an equation that describes the role of the thermal misbalance on the induced longitudinal flows is presented (see Equation \ref{AcousticWaveEq}). The analytical solution  expressed by Equation \ref{AcousticInt} sheds light on the features imposed by thermal misbalance on the induced longitudinal plasma motions due to shear Alfv\'en waves. The conclusions are summarized as 

\begin{enumerate}

\item The observed exponential bulk flow itself modifies the Alfv\'en-induced plasma motions. This bulk flow arises due to the fact that the thermal misbalance causes local increase/decrease in temperature, which leads to the increase in the pressure gradients that drive bulk flows. For the case where visco-resistivity is present, the same phenomena is experienced. But in the thermally active plasmas the flow may either travel in the same direction as the  Alfv\'en wave or travel opposite to it. The flow is co-directed with the Alfv\'en wave in the isentropically stable plasma ($C_\infty^2>C_0^2$ and $\tau_V>0$) and oppositely-directed in the isentropically unstable plasma  ($C_\infty^2<C_0^2$ and $\tau_V>0$).

\item { The Alfv{\'e}n wave induces the longitudinal plasma motions, which causes the thermal misbalance leading to additional plasma heating or cooling}. Subsequently, this heating/cooling introduces gas-dynamic pressure gradients driving longitudinal plasma motions that accelerate along the magnetic field. A simple mechanical analogue of this process is driven oscillations of a pendulum with a positive/negative friction. The external force with its own frequency moves a pendulum, but a positive/negative friction leads to the energy dissipation/gain that causes the oscillations to experience a phase shift. Based on this analogy, plasma particles act like a pendulum where the Alfv{\'e}n-wave ponderomotive force is an external force while the thermal misbalance plays the role of either positive or negative friction.

\item In particular, the amplitude and the phase shift of the oscillating part of longitudinal plasma motion induced by shear Alfv{\'e}n waves becomes frequency-dependent. This feature has been investigated analytically in application to coronal conditions. The heating  mechanism considered corresponds to the isentropically stable plasma \citep{Kolotkov_2020}. We note that the amplitude dependence is a monotonic function of the wave period (frequency) with maximum values in the high-frequency interval. At the same time, the negative phase shift (Figure \ref{fig-2}) is non-monotonic with a maximum observed at $945$ seconds {for the chosen heat-loss model and specific combination of the plasma parameters}.
\end{enumerate}

We have demonstrated that longitudinal plasma motions induced by shear Alfv\'en waves are affected by the thermal misbalance. This may contribute towards explaining the velocity shifts observed for propagating Alfv\'en waves, for instance in coronal holes, reported to be about $50\,\%$ of the amplitude \citep{2009A&A...501L..15B} and constraining the coronal-heating function. This provides a basis for taking a further step to shed light on how the torsional Alfv\'en wave acts as a driver for longitudinal plasma motions in the presence of thermal misbalance, which will provide insight on the influence of the heating and cooling effects on the interplay of the nonlinear forces connected with the torsional Alfv\'en wave in the context of coronal heating.

%


\begin{acks}
The study was supported in part by the Ministry of Education and Science of Russia by State assignment to educational and research institutions under Project No. FSSS-2020-0014 and No. 0023-2019-0003, and by RFBR, project number 20-32-90018. CHIANTI is a collaborative project involving George Mason University, the University of Michigan (USA), University of Cambridge (UK) and NASA Goddard Space Flight Center (USA).
\end{acks}

{\footnotesize\paragraph*{Disclosure of Potential Conflicts of Interest}
	The authors declare that they have no conflicts of interest. [Edit as appropriate.]
}

\bibliographystyle{spr-mp-sola}
\bibliography{refs}

\end{article} 
\end{document}